\documentstyle[nato,citesort]{crckapb}
\input epsf

\begin{opening}

\title{Free energy for liquids out of equilibrium}

\author{Antonio Scala}

\author{Francesco Sciortino}

\institute{
Dipartimento di Fisica ed Istituto Nazionale di Fisica della Materia,
Universit\`a degli studi di Roma ``La Sapienza'',P.le Aldo Moro 2,
I-00185, Roma, Italy\\ }

\end{opening}

\begin{document}

\begin{abstract}
Starting for the Stillinger and Weber expression for the free energy
of supercooled liquids, we extend the free energy to the case in which
two time scales separate and the system is in quasi-equilibrium. The
concept of an effective temperature, different from the kinetic
temperature, is naturally introduced. 
An example of hypotetical quasi-equilibrium phase diagram is 
presented for the case of SPC/E water.
\end{abstract}

\section{Introduction}

The study of the properties of undercooled liquids is a topic of
current research. Under the name of undercooled liquids we will
indicate liquids whose relaxations start to show a separation between
different timescales. The idea of considering the dynamics of
undercooled liquids as similar to the vibrations of a disordered solid
on short timescales, while attributing the slow dynamics to rare
change in the global structure (i.e. the atomic position of the
disordered solid assumed as the reference for short time vibrations)
is old~\cite{frenkel}; in particular, the fruitful reformulation of
Goldstein~\cite{goldstein} of these ideas in terms an energy landscape
has led to the introduction of a formalism~\cite{stillweb}
that is particularly suitable for the study of liquids via Monte Carlo
or Molecular Dynamics simulations~\cite{allentildesely}.

\begin{figure}[t]
\centering
\hbox to\hsize{\epsfxsize=0.5\hsize\hfil
\epsfbox{
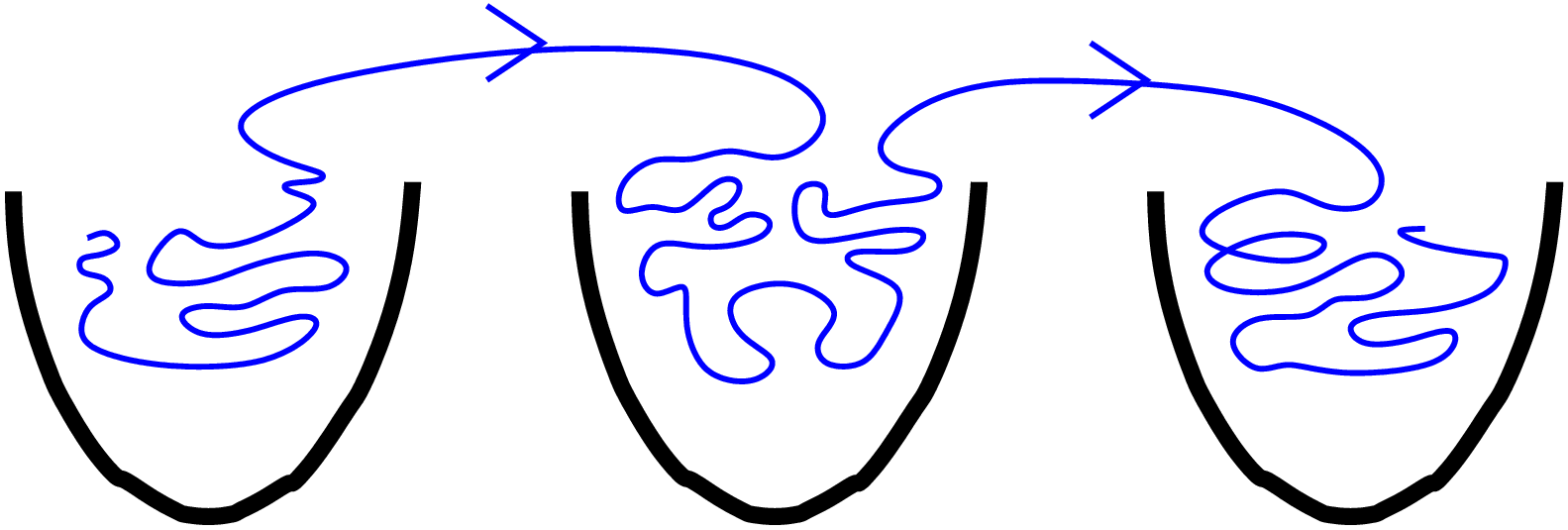
}\hfil}
\label{fig:goldstein}
\caption{Sketch of the undercooled liquids motion, according to
Goldstein~\protect\cite{goldstein}: vibrations in the basins of the
minima of the potential energy are interspersed by more unfrequent
change of basins.}
\end{figure}

In Sec.~\ref{sec:free.equil} we describe such formalism, showing that
it is particularly suitable to investigate the connections between the
statics and the dynamics of equilibrium systems. In
Sec.~\ref{sec:free.nonequil} we show that such formalism can be
extended to systems in quasi-equilibrium via the introduction of a
temperature for the slow degrees of freedom that can be measured both
via dynamics and via statics.  In Sec.~\ref{sec:SPCE.nonequil} we
speculate on the possibility of quasi-equilibrium, quench-rate
dependent critical points.

\section{Free energy for a supercooled liquid}
\label{sec:free.equil}

The phase space of a system can be in general partitioned in different
ways, giving rise to equivalent rewritings of the free
energy~\cite{palmer}.  In general, we will consider systems with an
hamiltonian ${\cal H}$ and therefore disregard the momentum space as
it can be trivially integrated out; we will therefore concentrate on
the $fN$-dimentional configuration space $W$, where $f$ is the number
of spatial degrees of freedoms of a particle and $N$ is the number of
particles.  What makes us prefer or introduce a particular partition
is its physical relevance ad interpretation.

\begin{figure}[t]
\centering
\hbox to\hsize{\epsfxsize=0.5\hsize\hfil
\epsfbox{
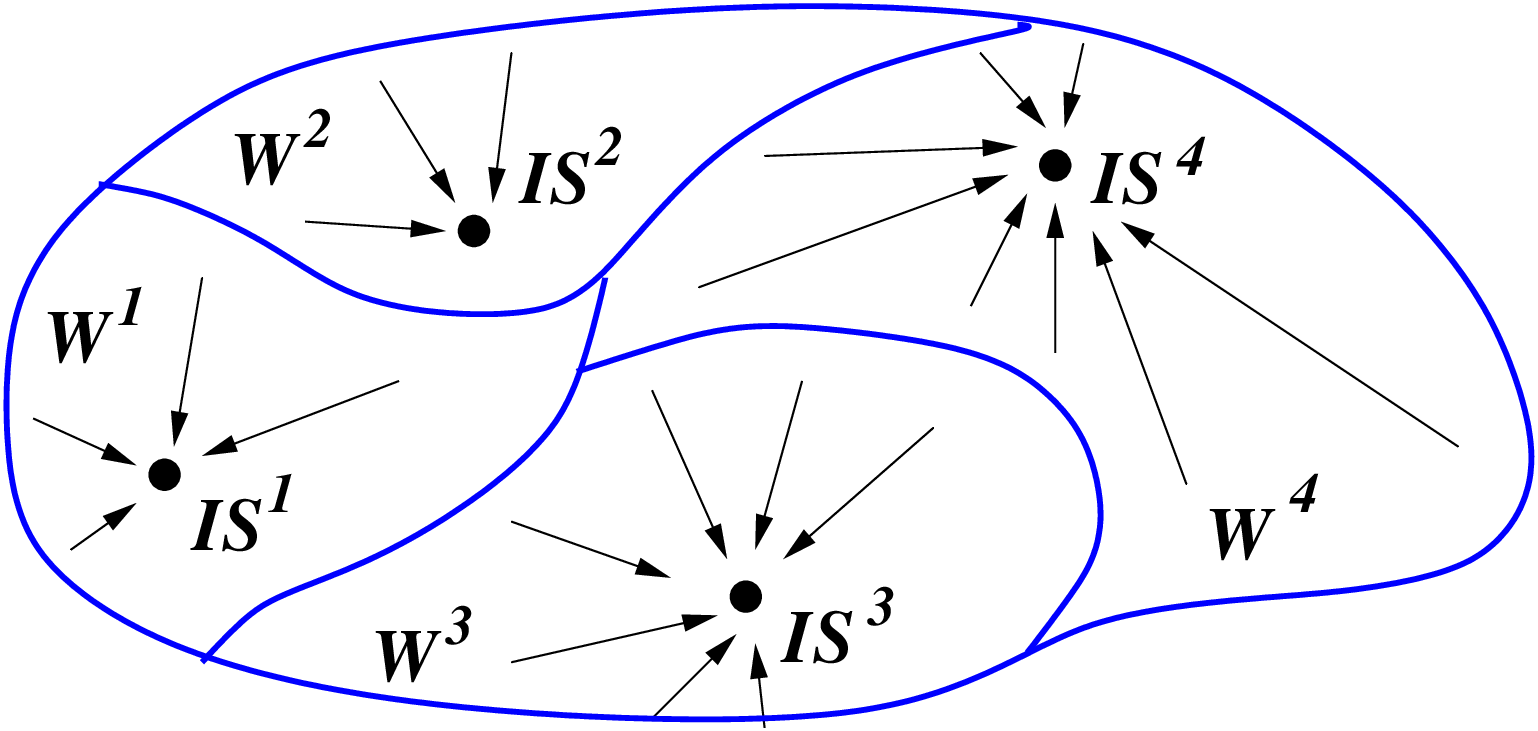
}\hfil}
\label{fig:stillpartition}
\caption{ Sketch of the Stillinger-Weber partition of the
configuration space $W$. The configurations indicated as $IS^i$ are
local minima of the potential energy; all the configurations of the
basin $W^i$ are associated to $IS^i$ via steepest descent.
}
\end{figure}

\begin{figure}[t]
\centering
\hbox to\hsize{\epsfxsize=0.5\hsize\hfil
\epsfbox{
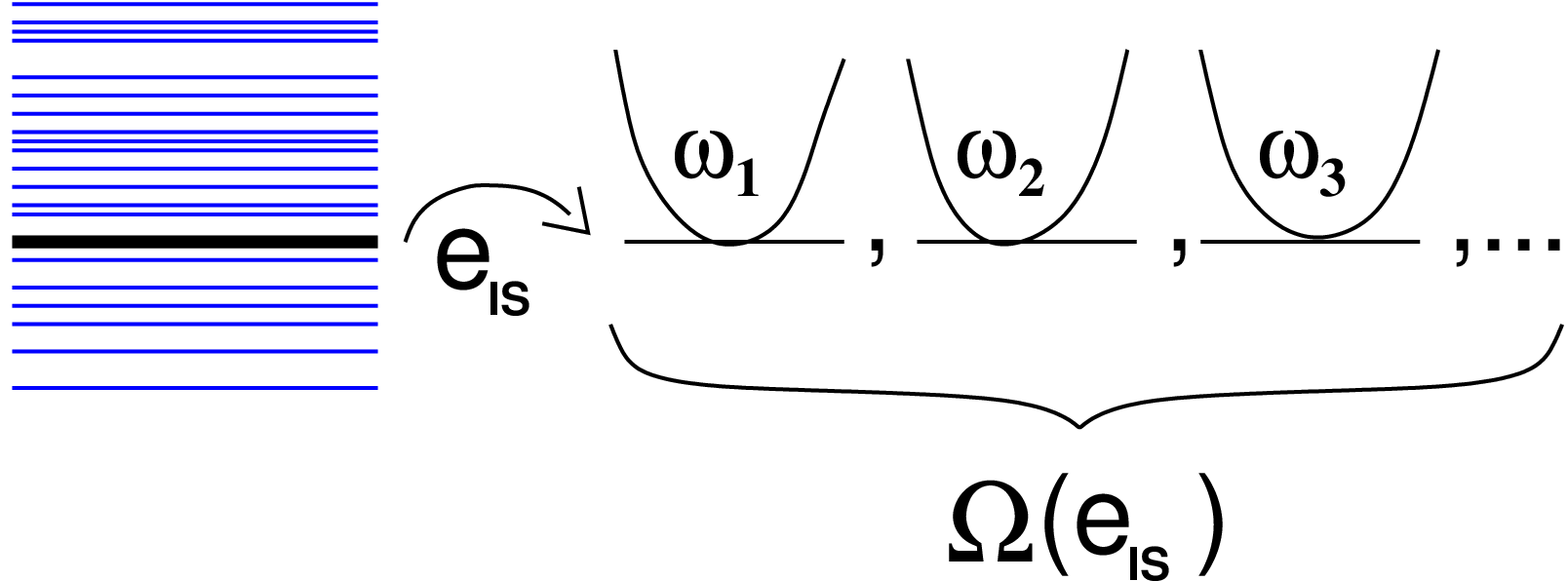
}\hfil}
\label{fig:eislevels}
\caption{The free energy of undercooled liquids can be expressed in
terms of the energy levels $e_{IS}$ of the inherent structures $IS$,
to their degeneracy $\Omega(e_{IS})$ and to the shape (in harmonic
approximation, to the curvature $\omega$) of the associated basins.}
\end{figure}

Stillinger and Weber introduced such a partition~\cite{stillweb} in
the spirit of the Goldstein picture of undercooled liquids dynamics
(see Fig.~\ref{fig:goldstein}). They associate to each configuration
the configuration reached following a steepest descent path to
minimize the potential energy. Such configuration is called an
Inherent Structure (IS) and corresponds to a local minimum of the
potential energy (Fig.~\ref{fig:stillpartition}). While to each IS we
can associate its potential energy $e_{IS}$, the converse is not true:
to each energy level $e_{IS}$ there will be associated a degeneracy
$\Omega(e_{IS})$ counting the number of local minima with energy
$e_{IS}$~\cite{caveat} (Fig.~\ref{fig:eislevels}). Following the
reasoning of Goldstein we can already argue that $e_{IS}$ is a good
candidate to be a slow variables of the system, as it is associated to
the global structure of a liquid and not to its vibrational
excitations. Defining the configurational entropy $s_{conf}=k_B
ln(\Omega(e_{IS}))$, we can write the Helmotz free energy of the
system at temperature $T$ and volume $V$ as~\cite{stillweb}

\begin{equation}
\label{eq:free.eq}
F(T,V) = e_{IS}(T,V) - T s_{conf}(e_{IS},V) + f_{vib}(e_{IS},T,V)
\end{equation}

\noindent
where $e_{IS}$ is the energy of the typical minima at $(T,V)$,
$s_{conf}$ for each fixed $V$ will depend implicitly on the
temperature via the T dependence of $e_{IS}$ and $f_{vib}$ is the
average free energy of the typical basins at $(T,V)$. At fixed $V$,
$f_{vib}$ will depend on the shape of the basins (we assume that the
shape of the basin can be parametrized by $e_{IS}$) and on the average
kinetic energy (therefore on T). The validity of such approximations
has been carefully checked and confirmed via numerical simulations of
different atomic and molecular model
systems~\cite{sciortTint,nature,speedysconf,heuer,srisconf,poolesconf}

For real systems $s_{conf}$ has been often approximated by the
difference of the entropy of the liquid and the entropy of some
reference solid phase~\cite{debenedetti}. For computer liquids we have
access to the IS configurations and can therefore extimate the
``true'' $s_{conf}$. In particular, we can use thermodynamic
integrations using the relation

\begin{equation}
\label{eq:deltasconf}
ds_{conf} = \frac{de_{IS}}{T} +\frac{df_{vib}}{T} = 
\frac{de_{IS}}{T} \left( 1+\frac{\partial f_{vib}}{\partial e_{IS}}\right)
\end{equation}

\noindent
for the differential of $s_{conf}$ obtained using the extremum
condition $\delta F = 0$. To calculate $F$, first an accurate estimate
of $e_{IS}$ must be obtained by minimizing the potential energy of
equilibrium configurations of the liquids. The estimate of $f_{vib}$
includes two steps: first, the harmonic contribution $f^{h}$ valid at
low temperatures is obtained calculating the normal mode frequencies
$\omega_i$ ($i=1..fN$~\cite{translations}) near the IS minima; the
exact calculation for a collection of harmonic oscillators gives us
the estimate~\cite{landaulifshitz}

\begin{equation}
\label{eq:fharm}
f^{h} = -k_B T << ln \left( \frac{h \omega}{2 \pi k_B T} \right) >>
\end{equation}

\noindent
where the average $<<..>>$~\cite{selfaveraging} is intended both on
the IS configurations and on all the $fN$ modes. For systems like
binary mixture Lennard Jones, $f^{h}$ is already a good estimate
for $f_{vib}$~\cite{sciortTint}. In the case of a molecular liquid,
the $SPC/E$ model for water~\cite{spce}, measurable deviations are
found~\cite{nature}. Still, deviations are small and can be estimated
as perturbative corrections to the ideal harmonic oscillator
behavior~\cite{feymann}. In practice, what it is found is that heating
up IS's one can estimate the anharmonic corrections $e_{anharm} =
aT^2+bT^3+...$ to the ideal harmonic behavior $e=e_{IS}+(f/2) k_B T$.

\begin{figure}[t]
\centering
\hbox to\hsize{\epsfxsize=0.5\hsize\hfil
\epsfbox{
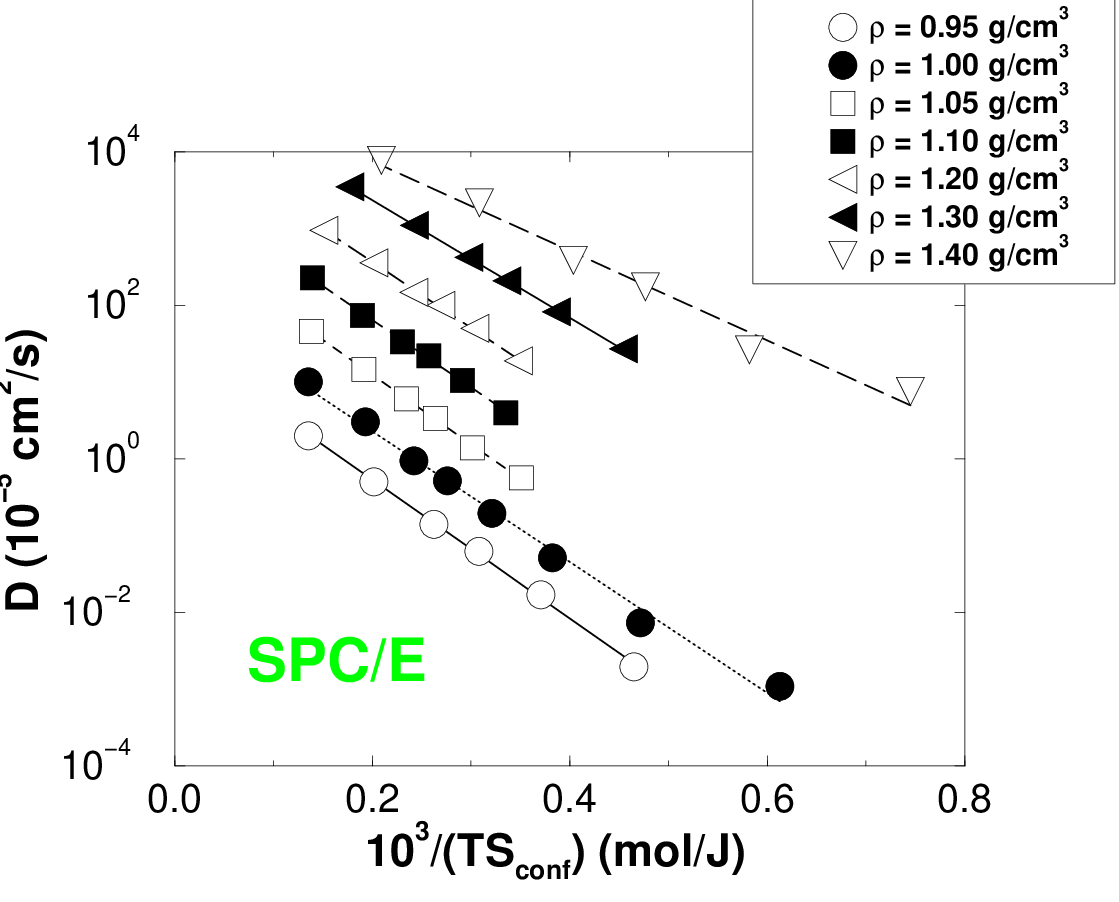
}\hfil}
\label{fig:adamgibbsSPCE}
\caption{Adam-Gibbs plot of diffusivity $D$ versus
$1/Ts_{conf}$~\protect\cite{nature}. The plot shows that there is a
strong correspondence among the dynamics and the statics of
undercooled liquids~\protect\cite{emiliaINM,angelani}. For the sake of
clarity, the values of $D$ have been shifted by multiplying by powers
of $5$ from $5^0$(the lowermost curve $\rho=0.95 g/cm^3$) to $5^6$
(the uppermost curve $\rho=1.40 g/cm^3$)}
\end{figure}

Therefore, once $e_{is}$ and $f_{vib}$ have been estimated, we can
recontruct $s_{conf}$ by thermodynamic integrations. The behavior of
$s_{conf}$ is related to the possibility of an ideal glass
transition~\cite{debenedetti}. If $s_{conf}$ becomes zero at some
finite temperature $T_K$ (the so-called Kauzmann
temperature~\cite{kauzmann}), this means that the liquid is in a
metastable state, trapped in the basin of a single minima; therefore
diffusivity is zero and the thermal vibrations around the disordered
structure (the IS of the minimum) are the only sort of motion taking
place. A precise functional dependence has been proposed by Adam and
Gibbs~\cite{adamgibbs} among $s_{conf}$ (a {\it static} quantity) and
the diffucion coefficient $D$ (a {\it dynamic} quantity):

\begin{equation}
\label{eq:adamgibbs}
D \propto e^{-\frac{A}{T s_{conf}}}
\end{equation}

where $A$ is almost constant with $T$~\cite{adamgibbs,corezzi}. Up to
now, the Adam-Gibbs relation~(Eq.~\ref{eq:adamgibbs}) has always been
found to be consistent with simulation
results~\cite{nature,srisconf,poolesconf,speedysconf} (see
Fig.~\ref{fig:adamgibbsSPCE}).

\section{Free energy for quasi-equilibrium states}
\label{sec:free.nonequil}

\begin{figure}[t]
\centering
\hbox to\hsize{\epsfxsize=0.5\hsize\hfil
\epsfbox{
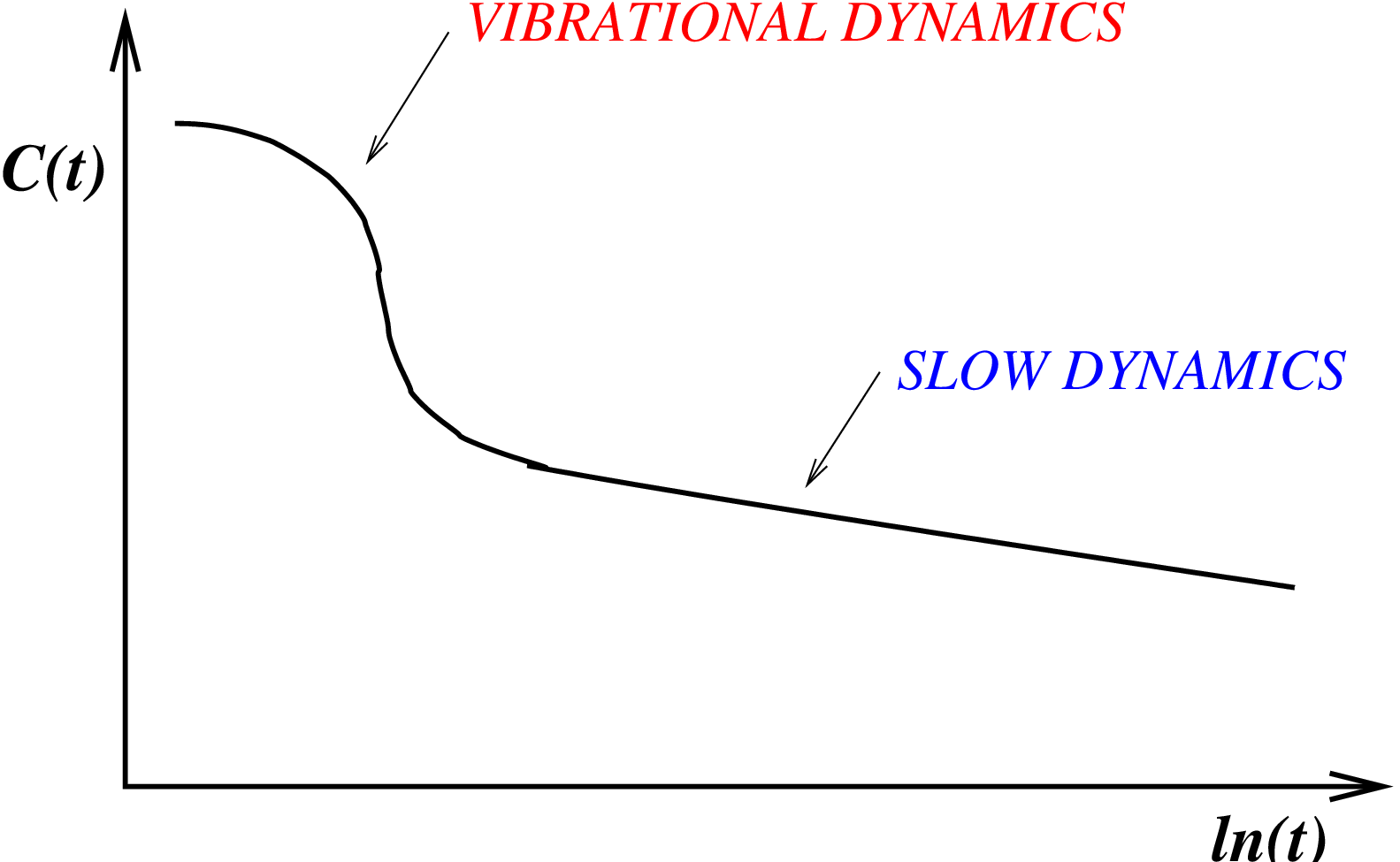
}\hfil}
(a)
\hbox to\hsize{\epsfxsize=0.5\hsize\hfil
\epsfbox{
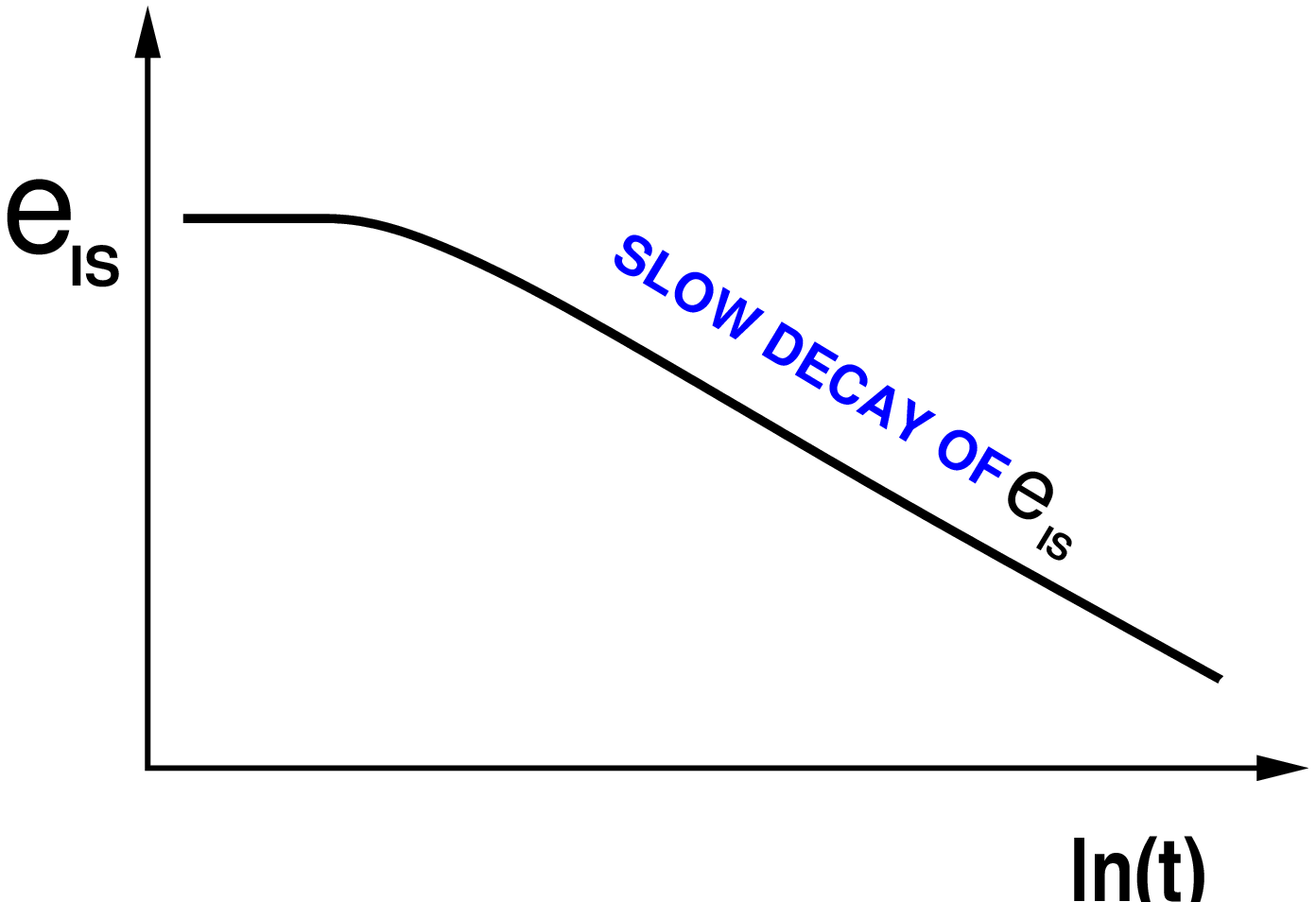
}\hfil}
(b)
\label{fig:slowcorr}
\caption{(a) The typical decay of a correlation function in
undercooled liquids. The fast relaxation corresponds to the
vibrational dynamics and can be treated in the harmonic approximation
using the istantaneous normal mode approach
(b) Typical decay of $e_{IS}$ after a deep quench. A slow logarithmic
law is compatible with numerical
data~\protect\cite{sciortTint,ruoccoSLOW}.}

\end{figure}

\begin{figure}[t]
\centering
\hbox to\hsize{\epsfxsize=0.5\hsize\hfil
\epsfbox{
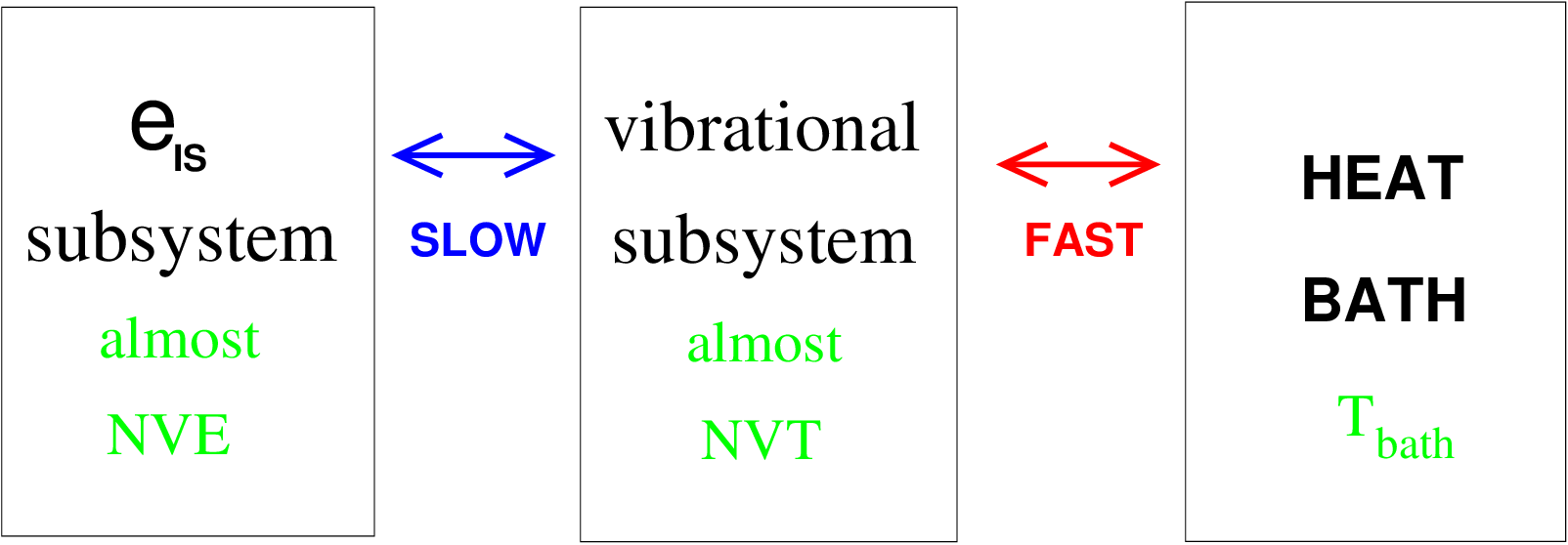
}\hfil}
\label{fig:slowbath}
\caption{Quenching the system (i.e. lowering rapidly the external
temperature $T_{bath}$ of the heath bath), vibrational degrees of
freedom come rapidly at a {\it canonical} (NVT) equilibrium with the
bath, while the subsystem corresponding to structural changes (the
$IS$'s) relaxes slowly and can be treated as an {\it almost
microcanonical} (NVE) system in which $e_{IS}$ does not change over
the characteristic vibrational timescales.}
\end{figure}

Attempts to construct thermodynamics for out-of-equilibrium states
have not yet come to a success; in particular, for glassy systems the
history of the system cannot in general be disregarded so that no
static theory with a small number of parameters can be
constructed~\cite{palmer}. In the case of undercooled systems, where a
separation between two different time scales becomes sharper and
sharper, it is possible to generalize in a straightforward way the
expression for the free energy. In the case of simulated liquids,
quenching at low $T$, the time evolution of $e_{IS}$ is slow, with a
decay compatible with a logarithmic or a very small power-law
decay~\cite{sciortTint}, while the short time dynamics has a much
faster timescale and is vibrational in character (see
Fig.~\ref{fig:slowcorr}). We can then treat our system as two
subsystems in contact: a vibrational subsystem, linked to the fast
vibrations in a basin, and the configurational subsytem of the local
minima $IS$'s, whose dynamics is reflected by the slow decay of
$e_{IS}$ (see Fig.~\ref{fig:slowbath}). On the timescales where the
vibrational subsystem comes in equilibrium with the external
heath-bath at temperature $T_{bath}$ (in this sense, the vibrational
subsystem is {\it canonical}~\cite{gibbs}), $e_{IS}$ is almost
constant (in this sense, the configurational subsystem is {\it almost
microcanonical}~\cite{gibbs}).

If we assume that the basins corresponding to a given $e_{IS}$ visited
in out-of equilibrium are the same (or have the same characteristics)
of the ones visited in equilibrium, we can use the expression of
$f_{vib}(e_{IS},T)$ obtained in equilibrium to take account of the
vibrational part of the free energy. Obviously, the vibrational
temperature will be equall to the external bath temperature
$T_{bath}$. The configurational part corresponds to the {\it almost
microcanonical} subsystem of the energy levels $e_{IS}$ together with
their degeneracies $\Omega(e_{IS})$. Now, we can write in general the
free energy of this subsystem as $e_{IS} - T_{int} s_{conf}(e_{IS})$
where $T_{int}$ (we will indicate $T_{int}$ as the {\it internal
temperature}) can be in general different from
$T_{bath}$~\cite{palmer}:

\begin{equation}
\label{eq:free.noneq}
F = e_{IS} - T_{int} s_{conf}(e_{IS}) + f_{vib}(e_{IS},T_{bath}).
\end{equation}

$T_{int}$ interpreted as a parameter that {\it ``optimizes''} the free
energy~\cite{jaynes} given the constraints that the energy of the
minima is $e_{IS}$ and that the vibrational subsystem has temperature
$T_{bath}$. Therefore, we can derive an expression of
$T_{int}$~\cite{franz,sciortTint} extremizing the generalized free
energy:

\begin{equation}
\label{eq:Tint}
\frac{\partial F}{\partial e_{IS}} = 0 \rightarrow  
T_{int}(e_{IS},T_{bath})=
\frac{ 1 + \frac{\partial f_{vib}}{\partial e_{IS}} }
{ \frac{\partial s_{conf}}{\partial e_{IS}} }.
\end{equation}

Up to now, we have a theoretical prescription for $T_{int}$ that is
self consistent with our writing of the generalized free energy. At
this point, two questions naturally arise: first, is it possible to
write the free energy in terms of experimentally measurable
quantities? And then, can we measure $T_{int}$?

Regarding the first problem, we recognize that already in the $30$'s
an approach to the construction of an out-of equilibrium free energy
for experimental glasses had been attempted by Davies and
Jones~\cite{daviesjones} in terms of ``frozen observables'',
i.e. quantities that mantain values typical of high temperatures after
quenching; a successful implementation of such ideas is found
in~\cite{speedy}, who uses the ``more'' concrete entalpies of shock
compressed hard-sphere liquids~\cite{stillHS} to calculate the
equilibrium configurational entropy.

\begin{figure}[t]
\centering
\hbox to\hsize{\epsfxsize=0.5\hsize\hfil
\epsfbox{
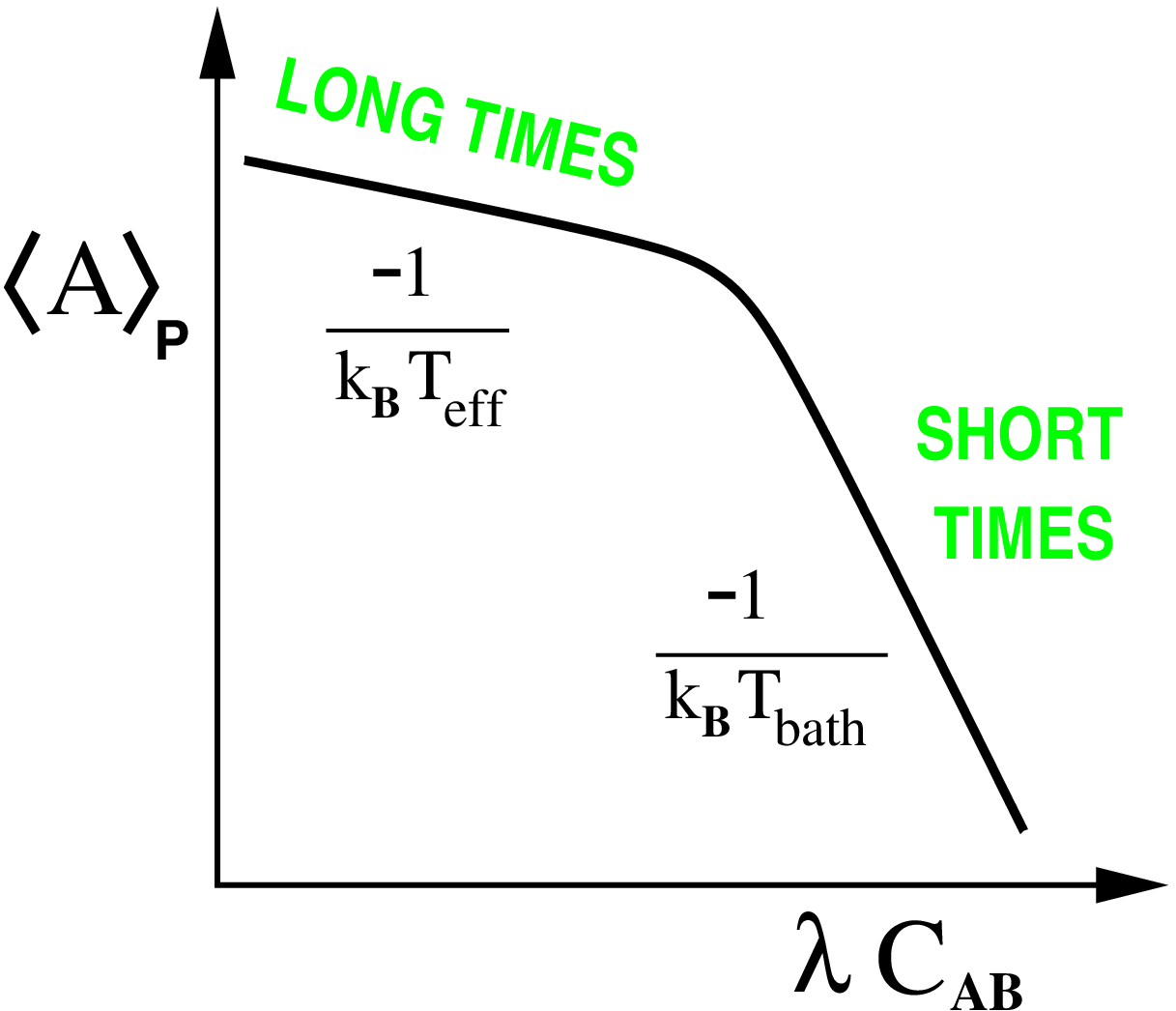
}\hfil}
\label{fig:fdt}
\caption{Fluctuation-dissipation plot of the response $<A>_P$ of the
system to a small perturbation versus the equilibrium, unperturbed
decay of the cross correlations $C_{AB}$ among observables $A$ and
$B$. According to Eq.~\protect\ref{eq:fdt}, the slope of the plot
gives a measure of the temperature of the system. While at short times
(i.e. high values of $C_{AB}$) the temperature of the system is in
equilibrium with the bath temperature, at long times (high values of
the response $<A>_P$) the system reveals that slow degrees of freedom
are at a temperature higher (lower slope of the plot) than the bath.}
\end{figure}

Regarding the measurement of $T_{int}$, we can resorts to dynamics
measuraments: it is in fact known both from theory~\cite{kucu},
numerical simulations~\cite{sciortTint} and
experiments~\cite{grigera}, that out of equilibrium systems can
exhibit at least two temperatures. A simple way to measure such
temperatures is through the application of linear response
theory~\cite{kubo}. Let's consider a pertubed hamiltonian
$H_P=H-\lambda B$, where $H$ is the unperturbed Hamiltonian, $B$ is a
generic observable and $\lambda$ is a small parameter. Linear response
theory predicts~\cite{hansenmcdonald} that, switching on the
perturbation at time $t=0$, we have

\begin{equation}
\label{eq:fdt}
<A(t)>_P = \frac{-\lambda}{k_B T}\left[C_{AB}(t)-C_{AB}(0)\right]
\end{equation}

\noindent
where $<A(t)>_P$ (the response of the system) is the time evolution of
the average of an observable $A$ {\it in presence} of the perturbation
and $C_{AB}(t)=<A(t)B(0)>$ is the correlation function (the decay of
fluctuations) between the observables $A$ and $B$ {\it in absence} of
the perturbation. Therefore, from the slope of a parametric plot of
$<A>_P$ versus $\lambda C_{AB}$, the temperature is measured from the
dynamics of the system (see Fig.~\ref{fig:fdt}).

While at short times such temperature coincides with $T_{bath}$, at
long times a temperature $T_{eff} \neq T_{bath}$ is measured in
quenching experiments. In the case of binary-mixture Lennard Jones
system, it has been verified that
$T_{eff}=T_{int}(e_{IS},T_{bath})$~\cite{sciortTint}. In general,
$T_{eff}$ tends to be higher than $T_{bath}$ and of the order of the
temperature at which the system is in equilibrium exploring basins of
depth $e_{IS}$. Somehow, the configurational part of the system
``remembers'' the temperature at which was originated.

\section{Out of equilibrium phase diagram for SPC/E water}
\label{sec:SPCE.nonequil}

Equation~\ref{eq:free.noneq} and~\ref{eq:Tint} allow to develop, using
the inherent structure formalism, an expression for the free energy
for disordered materials even in out-of-equilibrium states; for such
free energy, the ``history'' of the system is described by its
$e_{IS}$ value. In this section we present a preliminary application
to the case of SPC/E water.

The low temperature of SPC/E water has been object of a massive
computational investigations in order to investigate the slow dynamics
of molecular liquids~\cite{fstarrTMCT} and to check the liquid-liquid critical
point scenario~\cite{nature2ndCP}. In particular, the properties of the energy
landscape have been throughouly analized in order to find connections
between the dynamics and the statics of such a
system~\cite{nature,emiliaINM,landscapeSPCE}.

\begin{figure}[t]
\centering
\hbox to\hsize{\epsfxsize=0.5\hsize\hfil
\epsfbox{
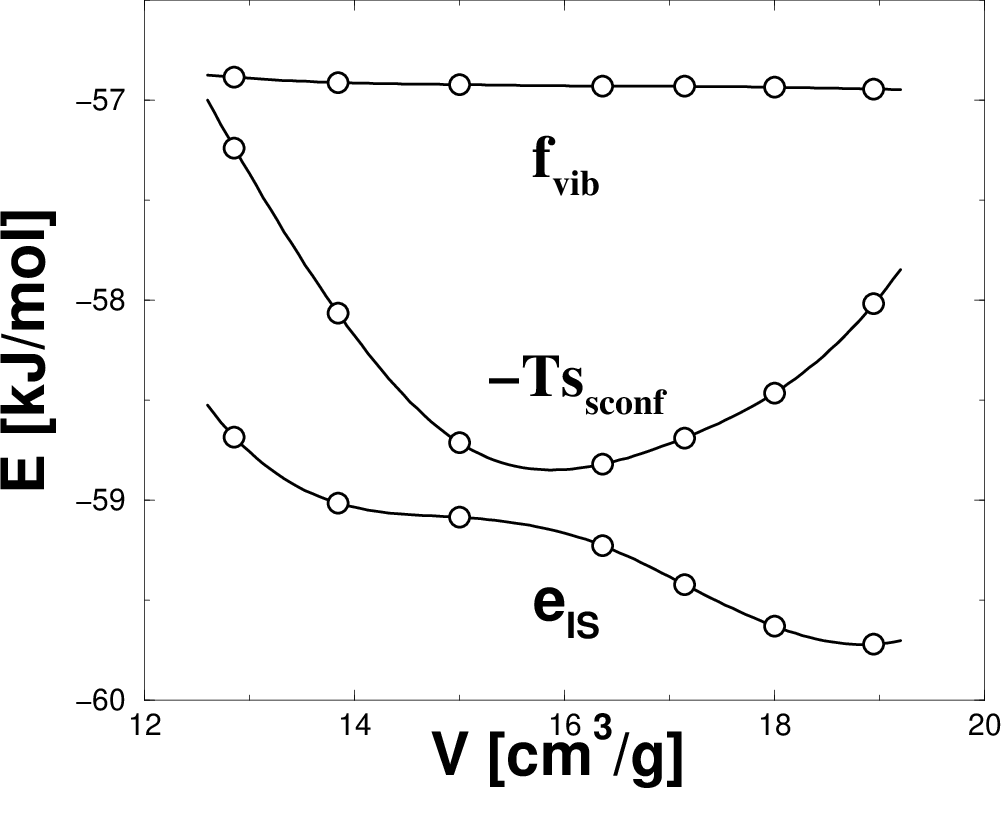
}\hfil}
\label{fig:SPCE210}
\caption{$e_{IS}$, $s_{conf}$ and $f_{vib}$ contributions to the free
energy of SPC/E water at $T=210$~$K$. The shape of $e_{IS}$ indicates
that the system would phase separate if it was not for the strong
$-Ts_{conf}$ term which avoids the change of concavity in the free
energy. For SPC/E water, $f_{vib}$ is pratically irrelevant to change
the concavity of the free energy.}
\end{figure}

The profile of the energy of SPC/E versus the volume develops a change
in concavity already in the region numerically accessible (see
Fig.~\ref{fig:SPCE210}), suggesting the possibility of a phase
transition hindered by the entropic terms. Low temperature
extrapolations predict indeed a liquid-liquid critical point a
temperature of $\sim 130$~$K$ and a specific density of $\sim
1.1$~$g/cm^3$~\cite{freeSPCE}. Such a transition would be located
beyond the predicted Kauzmann line for SPCE water~\cite{nature} and
therefore unaccessible to equilibrium studies~\cite{note:Tk}.

To derive an expression for the free energy in the low temperature
region of SPC/E water, we select for simplicity the harmonic
approximation, where $f_{vib}(e_{IS},V,T_{bath}) \approx f^{h} = -k_B
T_{bath} << ln \left( \frac{h \omega(e_{IS},V)}{2 \pi k_B T_{bath}}
\right) >>$ (Eq.~\ref{eq:fharm}). We use the data for $s_{conf}$ at
$T=210$~\cite{nature} as reference entropy to obtain an expression for
$s_{conf}$:

\begin{equation}
\label{eq:sconfit}
s^{h}_{conf}(e_{IS},V) = s^{T=210}_{conf}(V) +
\int^{T(e_{IS},V)}_{210} \frac{de_{IS}+df^{h}}{T},
\end{equation}

\noindent
where $T(e_{IS},V)$ is the temperature at which a system at volume $V$
in equilibrium has $IS$'s of energy $e_{IS}$. We obtain then the
generic expression for the free energy in harmonic approximation:

\begin{equation}
\label{eq:Fharm}
F^{h}(e_{IS},T_{bath},V) = e_{IS} - T_{int}(e_{IS},T_{bath},V)
s^{h}_{conf}(e_{IS},V) + f^{h}(e_{IS},T_{bath},V).
\end{equation}

\begin{figure}[t]
\centering
\hbox to\hsize{\epsfxsize=0.5\hsize\hfil
\epsfbox{
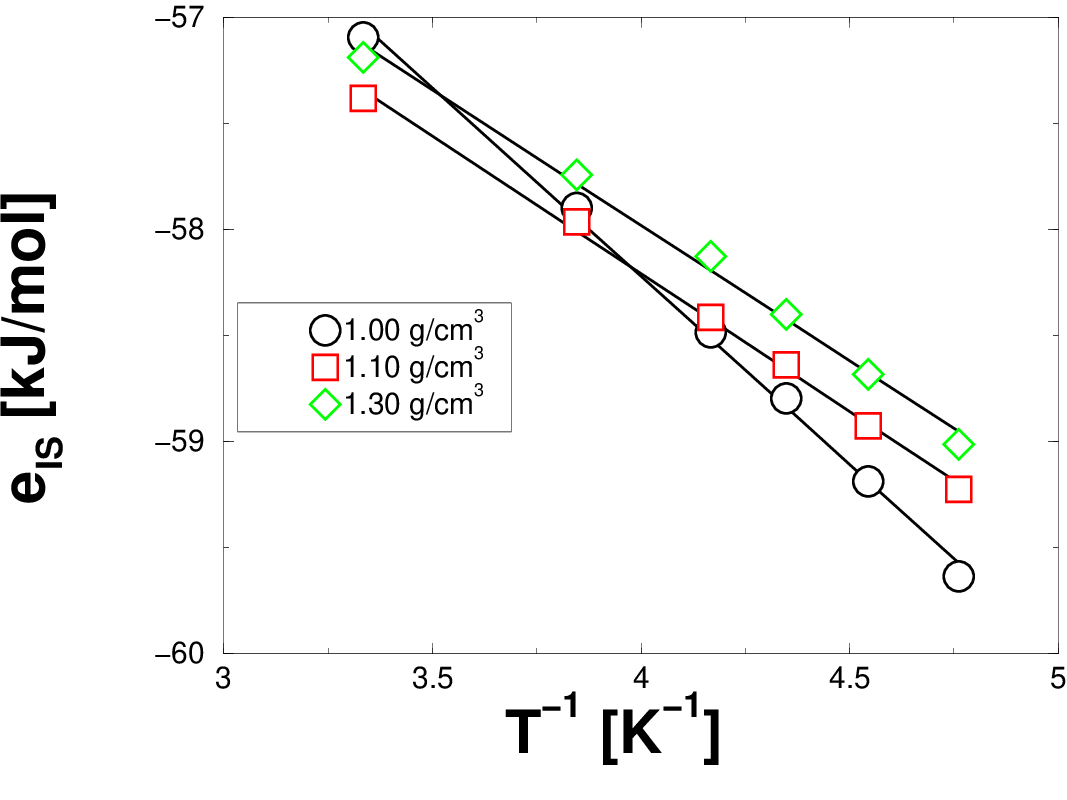
}\hfil}
(a)
\hbox to\hsize{\epsfxsize=0.5\hsize\hfil
\epsfbox{
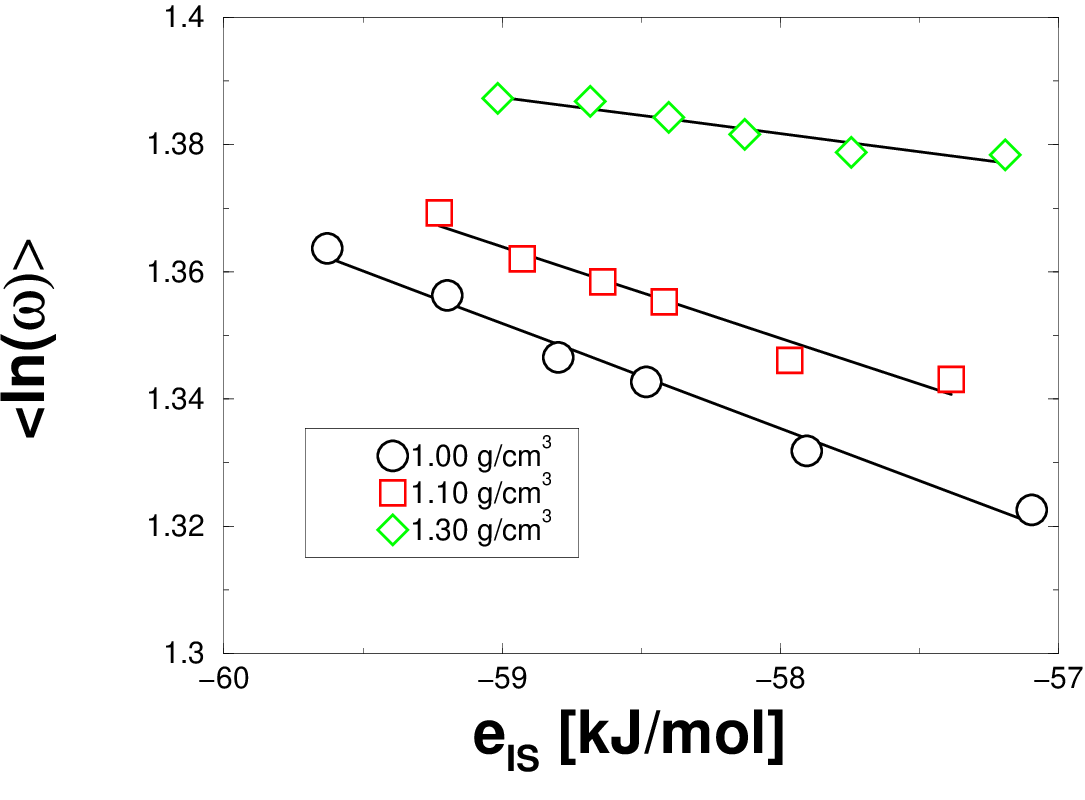
}\hfil}
(b)
\label{fig:linearSPCE}
\caption{(a) example of the linear dependence of $e_{IS}$ versus $1/T$
(b) the average logarithm of the basin's curvature is approximatively
linear in $e_{IS}$}
\end{figure}

First, we check the equilibrium phase diagram for $F^{h}$. No phase
transition is found in the physical region where $s_{conf} > 0$.
We then perform an ideal experiment: we consider the free energy of
state points in equilibrium at a temperature $T_{eq}$ after a sudden
quench at $T_{bath} = 100 K$ (well below the Kauzmann locus). After a
rapid quench, $e_{IS}$ has not time to change on the timescales of
vibrations, so that $e_{IS}$ and $s_{conf}$ remain to their
equilibrium values, while $f_{vib} \approx
f^{h}(\,e_{IS}(T_{eq}),\,T_{bath} = 100 K)$.

The value of the internal $T_{int}$ can be calculated from Eq.~\ref{eq:Tint}
and will be in general different from the temperature $T_{eq}$ at which the
quench has started. In the harmonic approximation, Eq.~\ref{eq:Tint} becomes
\begin{equation}
\label{eq:TintHarm}
T_{int}=T_{eq} \frac{1 - a\,T_{bath}}{1 - a\,T_{eq}}
\end{equation}
where $a= \frac{\partial << k_B ln \left[ \omega(e_{IS},V) \right]
>>}{\partial e_{IS}}$ and $T_{eq}$ is the temperature at which minima
of depth $e_{IS}$ are populated {\it in equilibrium}.

In the case of SPC/E water, as can be inferred from
Fig.~\ref{fig:linearSPCE}, $a$ is small (order $10^{-2}K^{-1}$) such that
$T_{int} \approx T_{eq}$, differing from the case of the binary
mixture Lennard-Jones~\cite{sciortTint} where $a$ is relevant and
$T_{int} < T_{eq}$ significantly.

Therefore, for SPC/E water in the harmonic approximation, it is not
possible to ``shift up'' the critical point above the Kauzmann
line. This is true for two reasons: first, in SPC/E water the
vibrational part $f_{vib}$ of the free energy has a very small
depenence both on $V$ and on $e_{IS}$ compared to the remaining
term. Therefore, if we ignore $f_{vib}$, the phase diagram stays
almost unchanged. The only way then to get a phase transition is to
``kill'' the entropic term $T_{int} s_{conf}$; but from
Eq.~\ref{eq:Tint} we see that, as $\partial f_{vib} /\partial e_{IS}
<< 1$, $T_{int} \approx T_{eq}$ where $T_{eq}$ will be approximatively
equal to the temperature at which the quench is started as the change
of $e_{IS}$ is logarithmically slow. So, for systems where $f_{vib}$
does not vary much with $e_{IS}$ (i.e. the shape of the basins does
not vary much), the temperature of the slow degrees of freedom of the
systems after a quench stays almost the same as the starting
temperature of the quench. In order to decrease drastically the
internal temperature (and therefore decrease the entropic term of the
free energy hindering eventual critical points), one has to resort is
a system with a strong dependence of the shape of the basins with
$e_{IS}$. In the extreme case $\partial f_{vib} /\partial e_{IS} >>
1$, we would have the ideal situation $T_{int} \approx T_{bath}$. The
possibility of the appearence of critical points in out-of-equilibrium
situation is a new issue that deserves to be further addressed and
investigated.

\section{Acknowledgments}

We thank Robin Speedy for stimulating discussions and acknowledge
partial support from INFM-PRA-HOP, INFM {\em Iniziativa Calcolo
Parallelo} and MURST-PRIN98.


\begin{thebibliography}{99}
\bibitem{caveat} Note that a rigorous analysis requires the definition
of a density of states $d(e_{IS})$ such that $d(e_{IS})\Delta$ are the
number of $IS$ with energy between $e_{IS}$ and
$e_{IS}+\Delta$~\protect\cite{khinchin}.
\bibitem{translations} 
Note that there are three spurious zero frequency modes corresponding
to the translation of the system.
\bibitem{selfaveraging} 
Note that for the derivation of the free energy of the equilibrium
liquid we are always assuming that self averaging for all
observables. This is not in general the case for systems with complex
landscapes at low temperatures~\cite{parisiSGTAB}
\end{thebibliography}
\end{document}